\documentclass[vecphys]{svmult}

\usepackage{makeidx}         
\usepackage{graphicx}        
                             
\usepackage{epsfig}
\usepackage{multicol}        
\usepackage[bottom]{footmisc}

\makeindex             

\begin{document}

\title*{Metallicity and mean age across M33}

\author{M.-R.L. Cioni\inst{1,3}, M. Irwin\inst{2}, A.M.N Ferguson\inst{3},
A. McConnachie\inst{4}, B.C. Conn\inst{5}, A. Huxor\inst{6}, R. Ibata\inst{7}, 
G. Lewis\inst{8}\and N. Tanvir\inst{9}} 

\authorrunning{Cioni, Irwin, Ferguson et al.}

\institute{Centre for Astrophysics Research, University of
  Hertfordshire, Hatfield, AL10 9AB, UK, \texttt{M.Cioni@herts.ac.uk}
\and Institute for Astronomy, University of Cambridge, Madingley Road,
  Cambridge CB3 0HA, UK, \texttt{mike@ast.cam.ac.uk}
\and SUPA, School of Physics, University of Edinburgh, IfA,
  Blackford Hill, Edinburgh EH9 3HJ, UK, \texttt{ferguson@roe.ac.uk}
\and Department of Physics \& Astronomy, University fo Victoria, PO
  Box, 3055, STN CSC, Victoria BC, V8W 3P6 Canada, \texttt{alan@uvic.ca}
\and European Southern Observatory, Alonso de Cordova 3107, Vitacura,
  Santiago, Chile, \texttt{bconn@eso.org}
\and Department of Physics, University of Bristol, Tyndall Avenue,
  Bristol BS8 1TL, UK, \texttt{Avon.Huxor@bristol.ac.uk}
\and Observatorie de Strasbourg, 11 rue de l'Universit\'{e}, F-67000
  Strasbourg, France, \texttt{ibata@astro.u-strasbg.fr}
\and Institute of Astronomy, School of Physics, A29, University of
  Sydney, NSW 2006, Australia, \texttt{gfl@physics.usyd.edu.au}
\and Department of Physics and Astronomy, University of Leicester,
  Leicester LE1 7RH, UK, \texttt{nrt3@star.le.ac.uk}}

\maketitle

New wide-field near-infrared (NIR) imaging observations of M33 were
obtained from UKIRT. These show a large population of intermediate-age
stars considerably improving on previous NIR data. The spatial
distribution of super giant stars, carbon-rich (C-rich or C stars) and
oxygen-rich (O-rich or M stars) asymptotic giant branch (AGB) stars
distinguished from the NIR colour-magnitude diagram (CMD) have been
studied as well as the C/M ratio. The $K_s$ magnitude distribution has
been interpreted using theoretical models to derive the mean age and
the mean metallicity across M33.

\section{Introduction}
\label{intro}
M33 is the third brightest Local Group (LG) member. It is of Sc II-III type,
thus intermediate between large spirals and dwarf irregular
galaxies. It has a nucleus, a disk and a halo and its stellar population
exhibits both a metallicity and age gradient (\cite{vdb}).

NIR observations of the stellar content of M33 began with
\cite{mcl}. While looking for a signature of a bulge component, which
was not found, they observed numerous intermediate-age stars in the
central $7.6^{\prime}$ of the galaxy down to $K\sim17-18$.  Several
years later (\cite{ste}) reaching $K\sim22$ detected young,
intermediate-age and old stars in the central
$22^{\prime\prime}$. Wide-field relatively shallow observations
($K_s\sim16$) by \cite{blo} claimed the existence of
arcs of metal poor C stars in the outer parts of M33.

\section{Observations, analysis \& results}
\label{obs}
New NIR observations of M33 have been obtained from UKIRT as part of a
programme to survey luminous red stars in LG galaxies (PI
Irwin). UKIRT data combine wide-field and good sensitivity improving
considerably on former studies. A mosaic of $4$ WFCAM tiles covering
$\sim3$ deg$^2$ was observed with an average seeing of
$1.07^{\prime\prime}\pm0.06^{\prime\prime}$. This allowed to reach
sources as faint as $K_s=18.32$ with S/N$=10$ including most
intermediate-age AGB stars. The data was dereddened assuming
$E(B-V)=0.07$ and using the \cite{gla} reddening law such that the
absorption in each wave band is $A_J=0.06$, $A_H=0.04$ and
$A_{K_s}=0.02$.

\begin{figure}
\centering
\includegraphics[height=6cm]{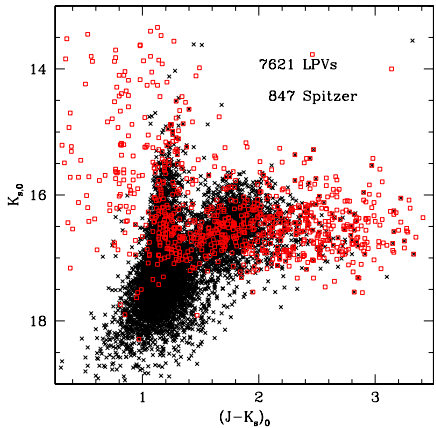}
\caption{CMD of NIR sources matched with LPVs
  candidates from Hartman et al. (\cite{har}; crosses) and Spitzer
  detections (squares). The TRGB is at $K_s=18.5$. C-rich
  AGB stars are redder than $J-K_s\sim1.5$ while O-rich stars are
  bluer. Stars observed by Spitzer with $J-K_s<1$ are likely supergiants.} 
\label{cmd}       
\end{figure}

The tip of the red giant branch (TRGB) is found at $K_s=18.15$. The
distribution of stars in the CMD ($J-K_s$, $K_s$) shows that
foreground stars, O-rich and C-rich AGB stars occupy clearly
distinct regions. Figure \ref{cmd} shows confirmed
long-period-variable (LPV) AGB stars from the cross-identification
with the variability study by \cite{har} as well as
obscured sources observed by Spitzer.

The spatial distribution of supergiant stars is clumpy and extends
asymmetrically to the NE while AGB stars trace a smoother distribution
with hints of the major galaxy spiral arms. The ratio between C- and
O-rich AGB stars (the C/M ratio) also outlines a ring-like feature and
suggests a metallicity spread of [Fe/H]$=0.6$ dex across the galaxy,
using the \cite{bat} calibration, in agreement with the results by
\cite{row}.
 
\subsection{Structure \& Extinction}

By subdividing the galaxy disk into $4$ concentric ellipses and $8$
sectors we investigated the orientation as well as the contribution by
differential extinction within the galaxy, if any. The peak of the
magnitude and colour differences from the mean, for C stars, describe
a sinusoidal pattern which indicates that stars in the NW of the
galaxy are fainter than stars in the SE of it. This sinusoid is
consistent with the distance moduli distribution derived by \cite{kim}
in $10$ different regions scattered within M33 suggesting that it is
almost entirely a structure rather then an extinction effect.

\subsection{Mean age and metallicity}

The $K_s$ magnitude distribution of stars within each sector of each
ellipse has been fitted with theoretical distributions spanning a
range of mean ages ($2-10.6$ Gyr) and mean metallicities
(Z$=0.0005-0.016$). The theoretical models used to create the
distributions are those by \cite{gir}. This method was first used by
\cite{cio} to investigate the stellar population of the Large
Magellanic Cloud.

Maps of the best fit mean metallicity versus mean age and combined
maps of best fit mean age and metallicity have been created separately
for C-rich and O-rich AGB stars. These show an inhomogeneous
distribution of age and metallicity. Note that relative values are
much more significant than absolute values which can be model
dependent. 

The very central region of the galaxy or regions around it appear
metal rich compared to the overall disk. In particular, C stars trace
a disk/halo population which is metal poor ([Fe/H]$\le -1.6$ dex) while
the nucleus and other regions around it are as metal rich as
[Fe/H]$\sim -1.2$ dex. There is an outer thick ring of stars on
average $6$ Gyr old or older. O-rich AGB stars also suggest an old
($6-8.5$ Gyr) and metal poor ([Fe/H]$\le 0.5$ dex) outer ring while
the centre of the galaxy is as young ($1-5$ Gyr) and metal rich
([Fe/H]$\ge 0.3$ dex). Although trends are the same for both C-rich
and O-rich AGB stars, the latter show higher metallicity values but
very similar ages.

\begin{figure*}
\epsfxsize=0.38\hsize \epsfbox{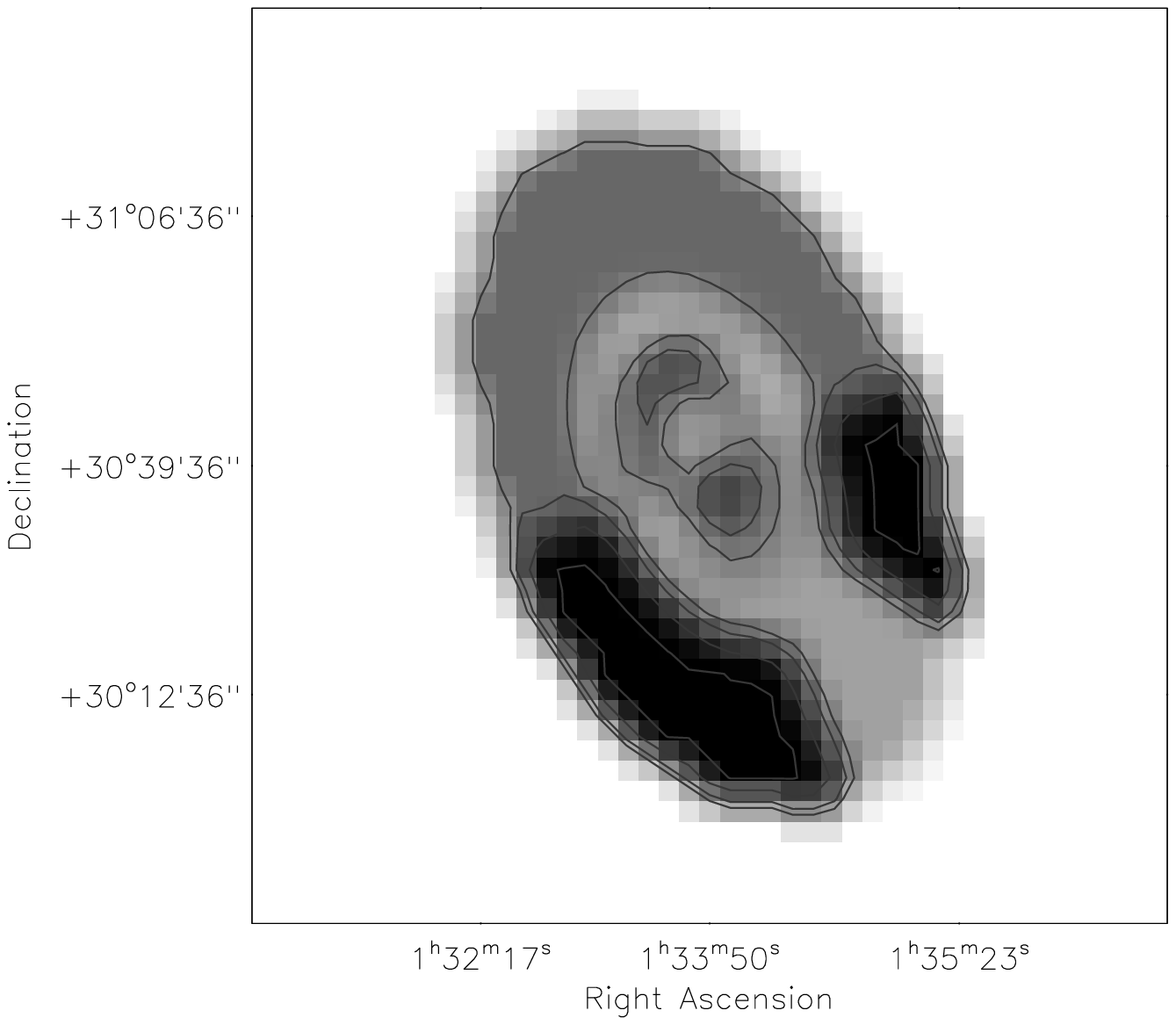}
\hspace{-1cm}
\epsfxsize=0.38\hsize \epsfbox{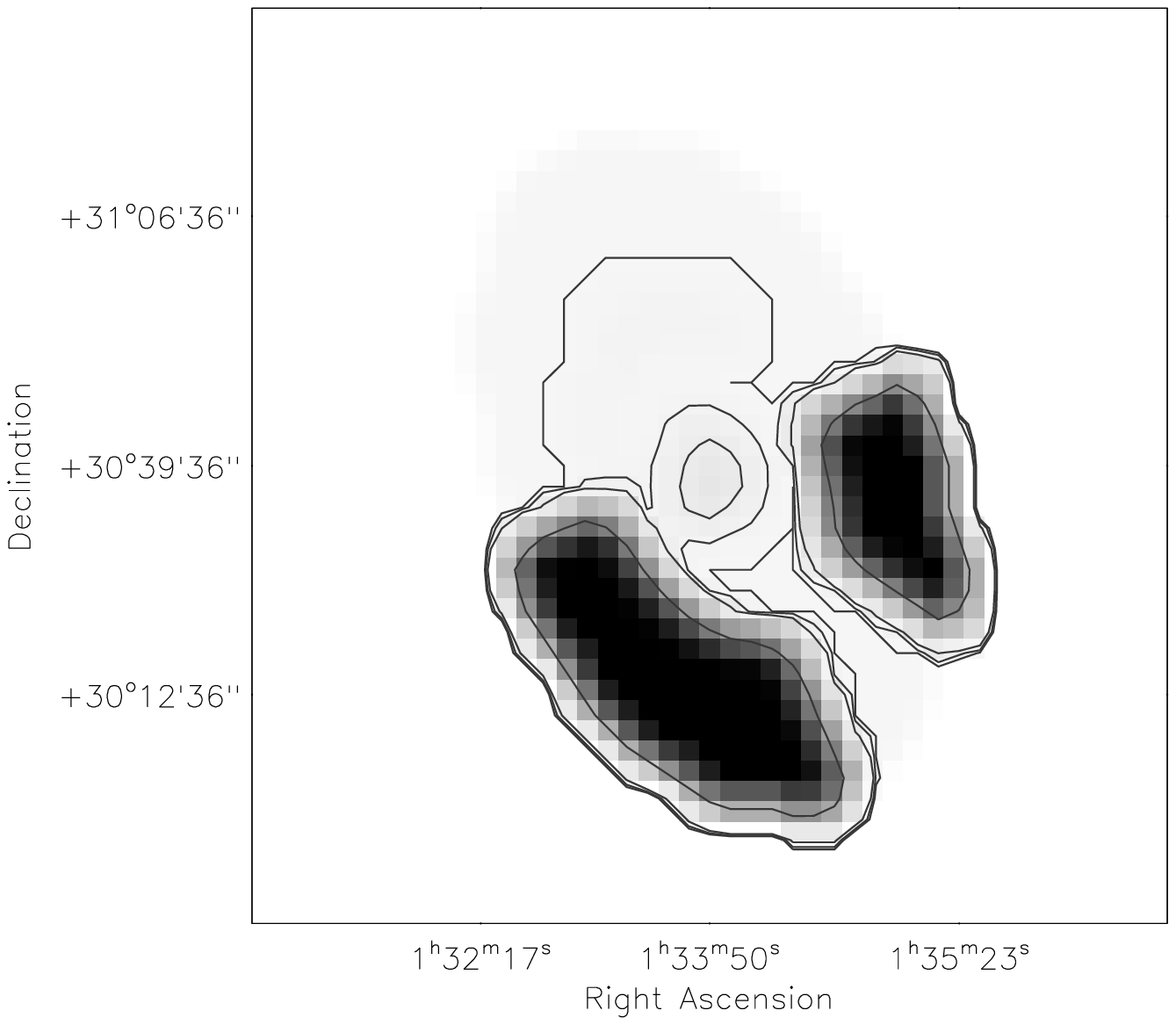}
\hspace{-1cm}
\epsfxsize=0.38\hsize \epsfbox{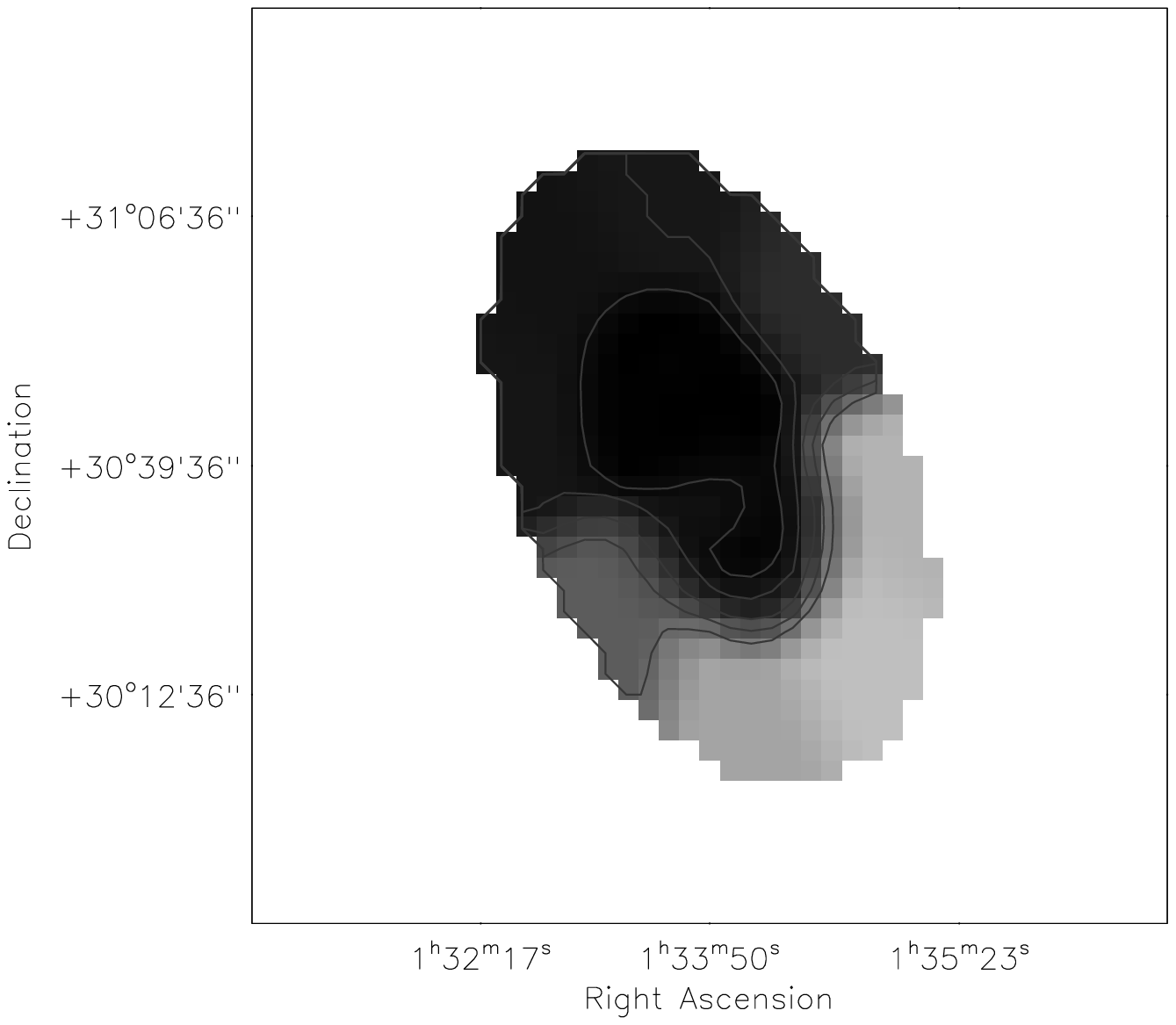} 

\epsfxsize=0.38\hsize \epsfbox{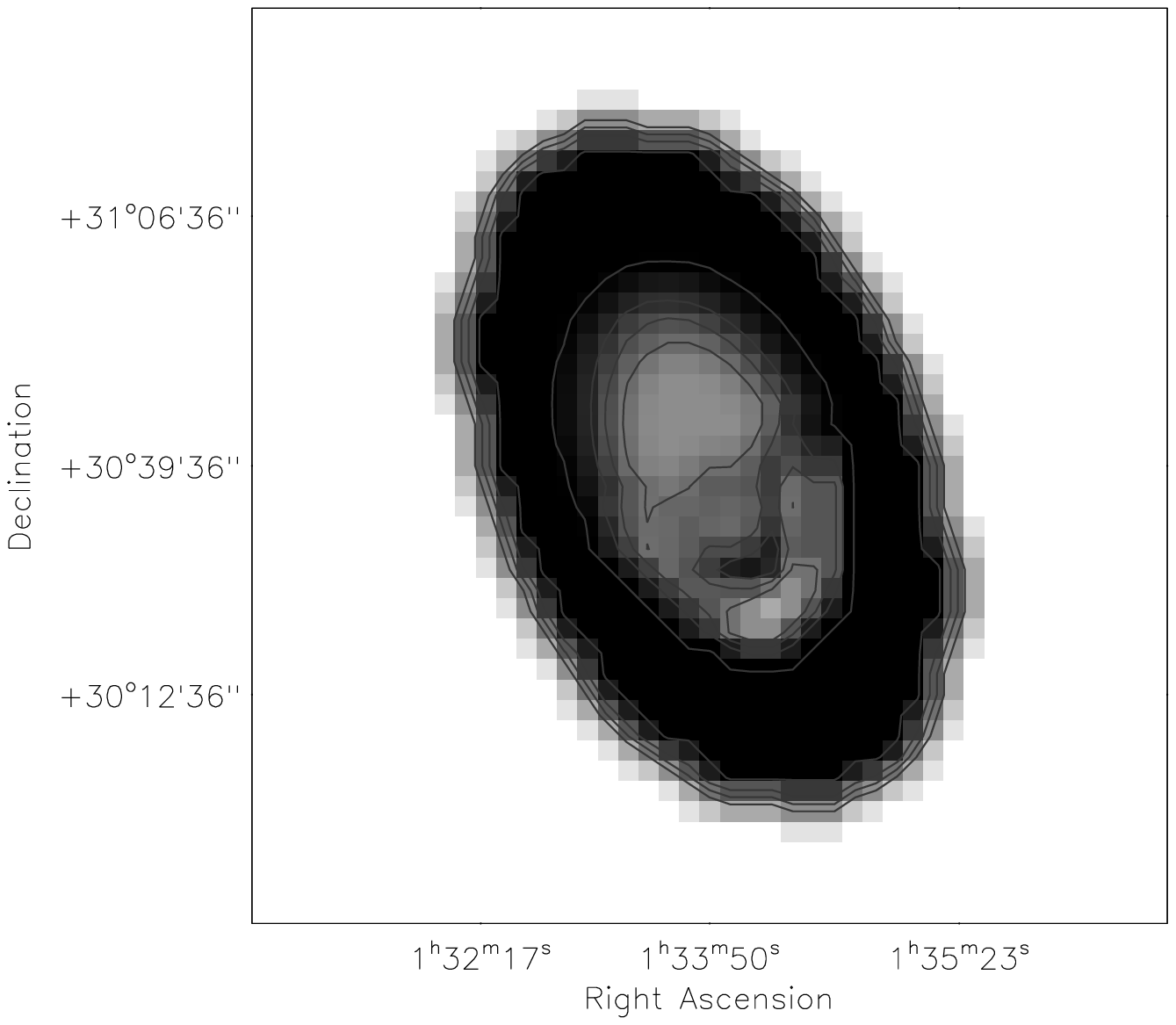}
\hspace{-1cm}
\epsfxsize=0.38\hsize \epsfbox{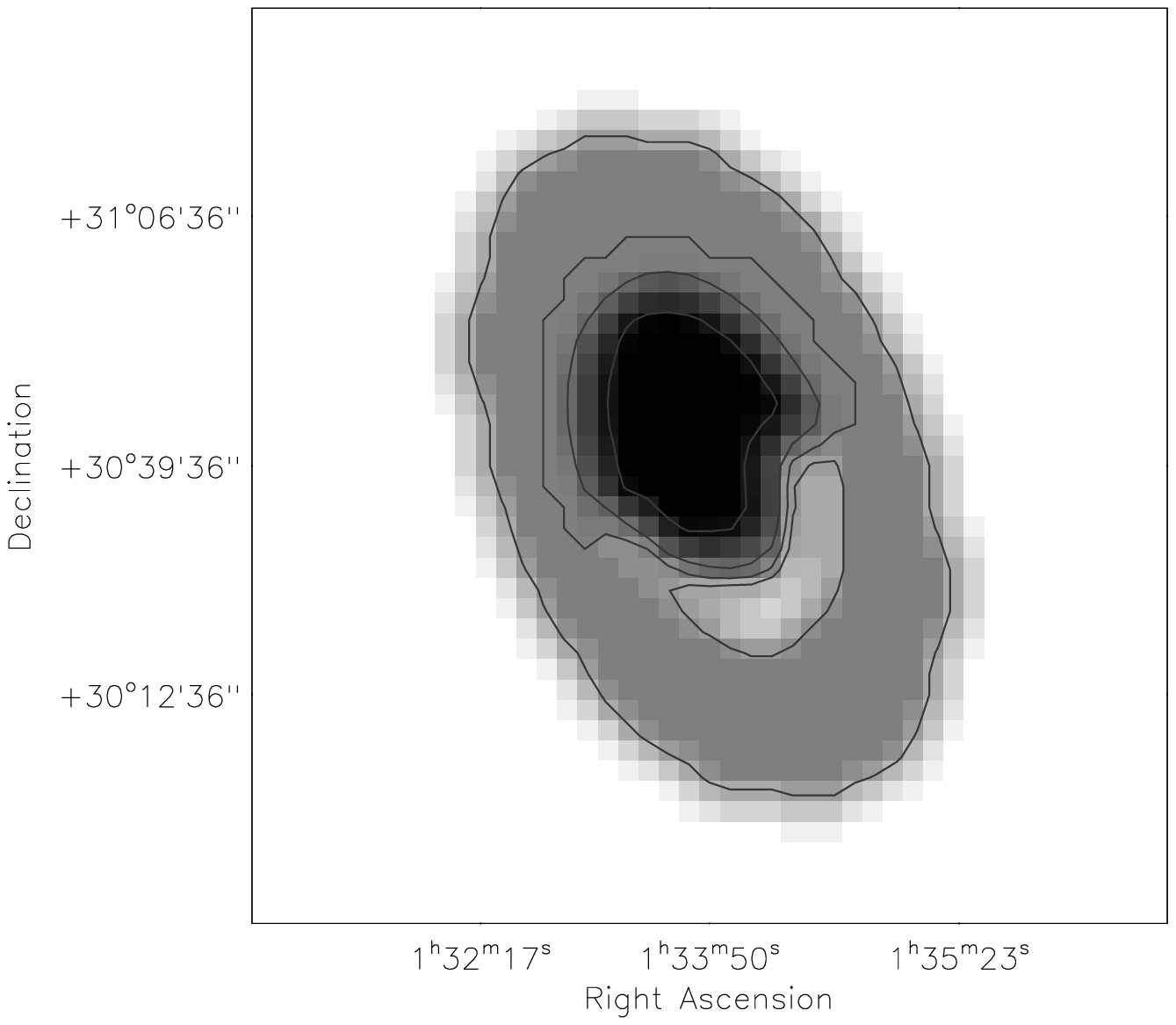}
\hspace{-1cm}
\epsfxsize=0.38\hsize \epsfbox{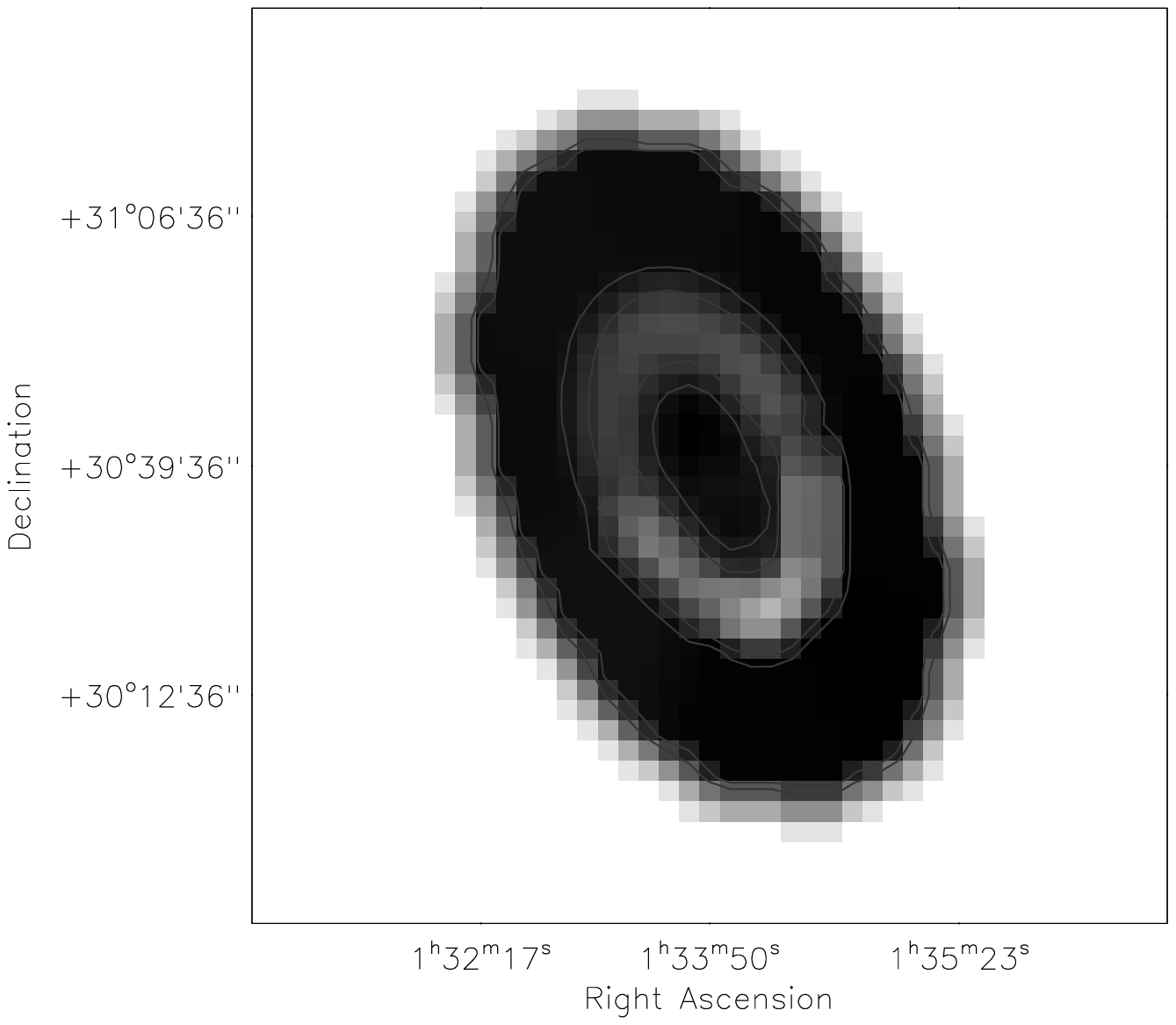} 
\caption{Spatial distribution of the mean age of the stellar
  population of M33 (top), of the metallicity (middle) and of the
  statistical probability that expresses the confidence level of the
  previous distributions. These distributions have been constructed
  from the comparison between the observed $K_s$ magnitude
  distribution of C-rich (top) and O-rich (bottom) AGB stars selected
  from the CMD. Darker regions correspond to higher numbers. North is
  up and East is left.}
\label{maps}
\end{figure*}

\section{Conclusions \& Future studies}
\label{con}

The $K_s$ method is an efficient way to constraint the parameters of a
galaxy stellar population using bright IR targets like AGB stars.  The
existence of a metallicity and age gradient throughout M33 is
confirmed. The C/M ratio, mean age and metallicity show much more
structure/substructures. The SE of the galaxy is closer to us; the
position angle of bright/faint stars is in the direction of the galaxy
warp. Many of the detected AGB stars are LPVs.

More galaxies have been observed within the same programme and will be
soon analysed. Theoretical models are already good enough to fit
entirely CMDs instead of just magnitude distributions. The study of
galaxies well beyond the LG has to wait next generation facilities
like JWST and E-ELT.

\index{paragraph}
%
%

%

\begin{thebibliography}{99.}
%
%
%


\bibitem{bat} P. Battinelli, S. Demers, A\&A \textbf{434}, 657 (2005)
\bibitem{blo} D.L. Block, K.C. Freeman, T.H. Jarret, et al. A\&A
  \textbf{425}, L437 (2004)
\bibitem{cio} M.-R.L. Cioni, L. Girardi, P. Marigo, H.J. Habing, A\&A,
  \textbf{448}, 77 (2006a)
\bibitem{gir} L. Girardi, A. Bressan, G. Bertelli, C. Chiosi, A\&AS,
  \textbf{141}, 371 (2000)
\bibitem{gla} I. Glass, M. Schultheis, MNRAS \textbf{345}, 39 (2003)
\bibitem{har} J.D. Hartman, D. Bersier, K.Z. Stanek, et al., MNRAS
  \textbf{371}, 1405 (2006)
\bibitem{kim} M. Kim, E. Kim, M.G. Lee, et al., AJ \textbf{123}, 244 (2002)
\bibitem{mcl} I.S. McLean, T. Liu, ApJ \textbf{456}, 499 (1996)
\bibitem{row} J.F. Rowe, H.B. Richer, J.P. Brewer, et al., AJ
  \textbf{129}, 729 (2005)
\bibitem{ste} A.W. Stephens, J.A. Frogel, AJ \textbf{124}, 2023 (2002)
\bibitem{vdb} S. van den Berg; \textit{The Galaxies of the Local
  Group}, Cambridge Astrophysics Series 35




\end{thebibliography}
%

\printindex
\end{document}